\documentclass[twocolumn,pra,aps,amssymb]{revtex4}
\usepackage{amsmath,amssymb,amsfonts}
\usepackage{graphicx,tikz,pgf,epsf}

\begin{document}
\title{Third-Party CNOT Attack on MDI QKD}
\author{Arpita Maitra}
\affiliation{Applied Statistics Unit,
Indian Statistical Institute,
203 B T Road, Kolkata 700 108, India,
Email: arpita76b@rediffmail.com
}
\begin{abstract}
In this letter, we concentrate on the very recently proposed Measurement Device
Independent Quantum Key Distribution (MDI QKD) protocol by Lo, Curty and 
Qi (PRL, 2012). We study how one can suitably present an eavesdropping 
strategy on MDI QKD, that is in the direction of the fundamental CNOT attack 
on BB84 protocol, though our approach is quite different. In this strategy, 
Eve will be able to know expected half of the secret bits communicated between 
Alice and Bob with certainty (probability 1) without introducing any error. 
Further, for the remaining bits, where Eve will only be able to predict the 
bit values as in random guess (with probability $\frac{1}{2}$), she will 
certainly find out whether her interaction induced an error in the secret bits 
between the communicating parties. Given the asymmetric nature of the CNOT 
attack, we also introduce Hadamard gates to present a symmetric version. 
Though our analysis does not refute the security claims in MDI QKD, adapting the
CNOT attack in this scenario requires nontrivial approach using entanglement
swapping.
\end{abstract}
\maketitle

\section{Introduction}
The idea of quantum key distribution was introduced by Bennet and Brassard, 
that is famous as the BB84 protocol~\cite{bb83,bb84}. 
Against BB84~\cite{bb84}, one of the most fundamental 
attack in this area is known as the CNOT attack that uses a CNOT gate. 
In this case, Eve can obtain complete information for the qubits sent in 
$Z$ basis without creating any disturbance. However, for the qubits sent in 
$X$ basis, Eve can not have any advantage and it also induces a disturbance as 
high as $\frac{1}{2}$. 

There are several variants of the traditional BB84 protocol that received 
attention in literature. The very recent proposals~\cite{side1,side2} are 
motivated from resistance against side channel attacks where they allow an 
untrusted party in the protocol. In particular,
to resist detector side channel attacks, measurement device independent 
quantum key distribution idea has been presented 
in~\cite{side2}. We will show how the fundamental idea of CNOT attack can be 
suitably modified to be accommodated in this scenario. As this proposal is very 
recent, to the best of our knowledge, such attack has not yet been studied.
The CNOT attack is inherently asymmetric. Thus, we exploit the Hadamard gate 
towards a symmetric version of this attack. 

In MDI QKD~\cite{side2}, Alice and Bob need not measure any qubit, and all the 
measurements are executed at Eve's end, an untrusted third-party. Thus, for 
eavesdropping strategies, it is natural to consider that Eve herself will try 
to gather information about the secret key while assisting Alice and Bob. That 
is why, this attack can be termed as third-party attack. While the idea 
of~\cite{side1} uses entanglement swapping~\cite{es} for building the protocol,
it is interesting to note that we exploit this for third-party CNOT attack
against MDI QKD~\cite{side2}. The application of entanglement swapping is 
evident in such protocols (either in design or in analysis) due to the 
involvement of the third-party.

Let us now present a few notations that we will be using.
By $BER_{AB}$ we denote the Bit Error Rate for the key bits between 
Alice and Bob. By $P_E$, we denote the Success Probability of Eve in 
correctly guessing the bit that Alice sent to Bob in form of a qubit. The
eavesdropping technique (that we present here) considers that Eve will
either get the complete information about the bit, i.e., $P_E = 1$ or
she will have no information at all other than the random guess, i.e.,
$P_E = \frac{1}{2}$. However, in the second case, Eve will have some other 
kind of information as follows.
By $\pi_E$, we denote the success probability of Eve in correctly guessing 
whether an error gets introduced during the communication between Alice and 
Bob. That is, in this case, Eve may not have any knowledge about the
value of the bit, but she exactly knows whether an error has occurred or
not during the communication between Alice and Bob, i.e., $\pi_E = 1$.

\section{CNOT attack on MDI QKD~\cite{side2}}
\label{three}
To understand this algorithm, we use Bell states. These are
two-qubit entangled states that can form orthogonal basis. The four 
Bell states can be written as 
$|\Phi^\pm\rangle = \frac{1}{\sqrt{2}}[|00\rangle \pm |11\rangle], \
|\Psi^\pm\rangle = \frac{1}{\sqrt{2}}[|01\rangle \pm |10\rangle].$
The untrusted third-party Eve measures the states received from Alice and Bob 
in this basis and informs the measurement result back to them. 
For eavesdropping purposes, we will also study some other measurements
by Eve on the qubits through which she will interact with the qubits sent by
Alice and Bob. For such purposes, based on the public discussion between Alice
and Bob, Eve will either measure in Bell basis or in computational basis,
i.e., $|00\rangle, |01\rangle, |10\rangle, |11\rangle$.
Before proceeding further, let us first explain MDI QKD~\cite{side2}.

\begin{enumerate}
\item Alice and Bob create random bit strings at their ends and encodes the 
bits in either $Z$ or $X$ basis randomly and send those to Eve.

\item Eve receives each pair of qubits (one from Alice and one from Bob) 
and measures them in Bell basis.
The detection results are publicly announced.

\item For the cases where the basis of Bob and Alice match
\begin{enumerate}
\item if the qubits of Alice and Bob are in $Z$ basis and the measurement 
results at Eve are $|\Psi^\pm\rangle$,
one of Alice or Bob has to flip the bit;

\item if the qubits of Alice and Bob are in $X$ basis and the measurement 
result at Eve is $|\Psi^-\rangle$ or $|\Phi^-\rangle$,
one of Alice or Bob has to flip the bit;
\end{enumerate}

\item Information reconciliation (using error correcting codes) and 
privacy amplification are performed by Alice and Bob on the remaining $n$ bits 
(let us call that the {\sf raw key}) to obtain $m$ shared key bits ({\sf final
key}).
\end{enumerate}
In the actual implementation, 
Eve can identify only two ($|\Psi^\pm\rangle$) of the four Bell states 
and that is claimed to be enough for the security proof to go 
through~\cite{side2}. Our analysis will also go through in a similar manner
in such a scenario.

We present the following table for understanding all the cases. When Alice and 
Bob generate qubits in different bases then those pairs of qubits are 
discarded and thus this is not shown in the table. 
\begin{center}
\begin{tabular}{|c|c|c|c|c|c|c|}
\hline
\multicolumn{2}{|c|}{Qubits sent by} & \multicolumn{4}{c|}{Probability (Eve's end)} & Flip\\
\hline
Alice & Bob & $|\Phi^+\rangle$ & $|\Phi^-\rangle$ & $|\Psi^+\rangle$ & $|\Psi^-\rangle$ & \ \\
\hline
\hline
$|0\rangle$ & $|0\rangle$ & $\frac{1}{2}$ & $\frac{1}{2}$ & 0 & 0 & No\\
$|0\rangle$ & $|1\rangle$ & 0 & 0 & $\frac{1}{2}$ & $\frac{1}{2}$ & Yes\\
$|1\rangle$ & $|0\rangle$ & 0 & 0 & $\frac{1}{2}$ & $\frac{1}{2}$ & Yes\\
$|1\rangle$ & $|1\rangle$ & $\frac{1}{2}$ & $\frac{1}{2}$ & 0 & 0 & No\\
\hline
$|+\rangle$ & $|+\rangle$ & $\frac{1}{2}$ & 0 & $\frac{1}{2}$ & 0 & No\\
$|+\rangle$ & $|-\rangle$ & 0 & $\frac{1}{2}$ & 0 & $\frac{1}{2}$ & Yes\\
$|-\rangle$ & $|+\rangle$ & 0 & $\frac{1}{2}$ & 0 & $\frac{1}{2}$ & Yes\\
$|-\rangle$ & $|-\rangle$ & $\frac{1}{2}$ & 0 & $\frac{1}{2}$ & 0 & No\\
\hline
\end{tabular}
\end{center}
\subsection{The CNOT attack}
\label{three1}
The eavesdropping model in this case is as follows, where the untrusted 
third-party Eve will try to obtain the information. Eve will take the 
qubits from
Alice and Bob and put each one of them in the control input of a CNOT gate and 
she will supply $|0\rangle$ in the target. The outputs corresponding to the 
control qubits of the CNOT gates will be measured in the Bell basis by Eve and 
the result will be communicated to Alice and Bob. Eve stores the output 
corresponding to the target in her quantum memory. Then Alice and Bob
will go for public discussion to announce their bases. Knowing these,
Eve will try to extract information from the outputs corresponding to
the target qubits of the CNOT gates.

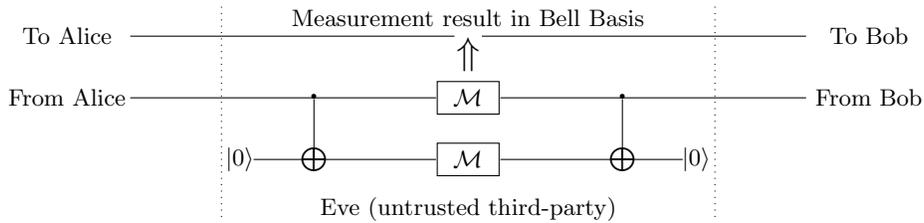
\begin{figure*}[htb]
\begin{center}
\begin{tikzpicture}[scale=0.82]
\node [inner sep=0pt,outer sep=0pt](p1) at (1,1) {};
\node [inner sep=0pt,outer sep=0pt](p2) at (4,1) {};
\node [inner sep=0pt,outer sep=0pt](p3) at (-1,2) {};
\node [inner sep=0pt,outer sep=0pt](p4) at (4,2) {};
\node [inner sep=0pt,outer sep=0pt](p5) at (2,1) {};
\node [inner sep=0pt,outer sep=0pt](p6) at (2,2) {};
\draw [-] (p1) -- (p2);
\draw [-] (p3) -- (p4);
\draw [-] (p5) -- (p6);
\node [inner sep=0pt,outer sep=0pt] (xor1) at (2,1) {$\bigoplus$};
\node [inner sep=0pt,outer sep=0pt] (dot1) at (2,2) {{\LARGE$\cdot$}};

\node [inner sep=0pt,outer sep=0pt](p1) at (5,1) {};
\node [inner sep=0pt,outer sep=0pt](p2) at (8,1) {};
\node [inner sep=0pt,outer sep=0pt](p3) at (5,2) {};
\node [inner sep=0pt,outer sep=0pt](p4) at (10,2) {};
\node [inner sep=0pt,outer sep=0pt](p5) at (7,1) {};
\node [inner sep=0pt,outer sep=0pt](p6) at (7,2) {};
\draw [-] (p1) -- (p2);
\draw [-] (p3) -- (p4);
\draw [-] (p5) -- (p6);
\node [inner sep=0pt,outer sep=0pt] (xor1) at (7,1) {$\bigoplus$};
\node [inner sep=0pt,outer sep=0pt] (dot1) at (7,2) {{\LARGE$\cdot$}};

\node [inner sep=0pt,outer sep=0pt] (t1) at (-2,2) {From Alice};
\node [inner sep=0pt,outer sep=0pt] (t2) at (0.8,1) {$|0\rangle$};
\node [inner sep=0pt,outer sep=0pt] (t2) at (8.2,1) {$|0\rangle$};
\node [inner sep=0pt,outer sep=0pt] (t2) at (11,2) {From Bob};
\node [inner sep=0pt,outer sep=0pt] (t1) at (-2,3) {To Alice};
\node [inner sep=0pt,outer sep=0pt] (t2) at (11,3) {To Bob};
\node [inner sep=0pt,outer sep=0pt] (t2) at (4.5,0.2) {Eve (untrusted third-party)};

\node[rectangle, draw=black, minimum height=2, minimum width=2] (H) at (4.5,1) {$ \ \cal{M} \ $};
\node[rectangle, draw=black, minimum height=2, minimum width=2] (H) at (4.5,2) {$ \ \cal{M} \ $};

\node [inner sep=0pt,outer sep=0pt](p1) at (0.5,3.5) {};
\node [inner sep=0pt,outer sep=0pt](p2) at (0.5,0) {};
\draw [dotted] (p1) -- (p2);
\node [inner sep=0pt,outer sep=0pt](p1) at (8.5,3.5) {};
\node [inner sep=0pt,outer sep=0pt](p2) at (8.5,0) {};
\draw [dotted] (p1) -- (p2);
\node [inner sep=0pt,outer sep=0pt](p1) at (-1,3) {};
\node [inner sep=0pt,outer sep=0pt](p2) at (4.3,3) {};
\draw [-] (p1) -- (p2);
\node [inner sep=0pt,outer sep=0pt](p1) at (10,3) {};
\node [inner sep=0pt,outer sep=0pt](p2) at (4.7,3) {};
\draw [-] (p1) -- (p2);
\node [inner sep=0pt,outer sep=0pt] (t1) at (4.5,2.7) {\Large{$\Uparrow$}};
\node [inner sep=0pt,outer sep=0pt] (t2) at (4.5,3.3) {Measurement result in Bell Basis};
\end{tikzpicture}
\caption{CNOT attack on MDI QKD}\label{fig6}
\end{center}
\end{figure*}
Consider that both Bob and Alice communicated in $Z$ basis. In such a case,
Eve will be able to copy these perfectly using CNOT gates without creating any
disturbance to the qubits sent by Alice and Bob. If the measurement
output at Eve is $|\Phi^\pm\rangle$, then the bits of Alice and Bob match.
Similarly, if the measurement output at Eve is $|\Psi^\pm\rangle$, then the 
bits of Alice and Bob do not match and one of them needs to toggle his/her bit.
Thus in this case, Eve will obtain all the information without creating
any disturbance. Note that, in this case, Eve will measure her target qubit
in computational basis, i.e., $|00\rangle, |01\rangle, |10\rangle, |11\rangle$. 

When Bob and Alice communicate in $X$ basis, then error is introduced by the
CNOT attack and the situation can be seen as an example of entanglement
swapping~\cite{es}. Let us explain one specific case here. The other cases 
will be similar. Consider that Alice and Bob both send $|+\rangle$.
Thus, after the application of CNOT gates by Eve, there will be entangled 
states
$\frac{|0_A 0_{E_1}\rangle + |1_A 1_{E_1}\rangle}{\sqrt{2}}$ and
$\frac{|0_B 0_{E_2}\rangle + |1_B 1_{E_2}\rangle}{\sqrt{2}}$ corresponding
to Alice and Bob respectively. Now the qubits corresponding to Alice and
Bob will be measured in Bell basis. One can see that 
$$\left(\frac{|0_A 0_{E_1}\rangle + |1_A 1_{E_1}\rangle}{\sqrt{2}} \right)
\otimes
\left( \frac{|0_B 0_{E_2}\rangle + |1_B 1_{E_2}\rangle}{\sqrt{2}} \right)$$ 
can be written as 
$\frac{1}{2}(
|\Phi_{AB}^+\rangle|\Phi_{E_1E_2}^+\rangle +
|\Phi_{AB}^-\rangle|\Phi_{E_1E_2}^-\rangle +
|\Psi_{AB}^+\rangle|\Psi_{E_1E_2}^+\rangle +
|\Psi_{AB}^-\rangle|\Psi_{E_1E_2}^-\rangle
).$

The correct measurement in this case is 
$|\Phi_{AB}^+\rangle$ or $|\Psi_{AB}^+\rangle$ that happens with probability
$\frac{1}{2}$ and in such a case after the bases of Alice and Bob are 
published, Eve will measure either 
$|\Phi_{E_1E_2}^+\rangle$ or
$|\Psi_{E_1E_2}^+\rangle$ and she will be able to know that no error  
has been introduced. However, if the measurement result becomes 
$|\Phi_{E_1E_2}^-\rangle$ or $|\Psi_{E_1E_2}^-\rangle$ (this happens with probability
$\frac{1}{2}$ too), then Eve knows that an error has been introduced, and
she will not be able to know the secret bit.
Similarly, we can analyse the other cases and get the following as in 
Table~\ref{tab1}.
\begin{table*}[htb]
\begin{center}
\begin{tabular}{|c|c|}
\hline
Alice, Bob & Eve (after CNOT attack)\\
\hline
$|+\rangle, |+\rangle$ &
$\frac{1}{2}\left(
|\Phi_{AB}^+\rangle|\Phi_{E_1E_2}^+\rangle +
|\Phi_{AB}^-\rangle|\Phi_{E_1E_2}^-\rangle +
|\Psi_{AB}^+\rangle|\Psi_{E_1E_2}^+\rangle +
|\Psi_{AB}^-\rangle|\Psi_{E_1E_2}^-\rangle
\right)$\\
\hline
$|+\rangle, |-\rangle$ &
$\frac{1}{2}\left(
|\Phi_{AB}^+\rangle|\Phi_{E_1E_2}^-\rangle +
|\Phi_{AB}^-\rangle|\Phi_{E_1E_2}^+\rangle -
|\Psi_{AB}^+\rangle|\Psi_{E_1E_2}^-\rangle -
|\Psi_{AB}^-\rangle|\Psi_{E_1E_2}^+\rangle
\right)$\\
\hline
$|-\rangle, |+\rangle$ &
$\frac{1}{2}\left(
|\Phi_{AB}^+\rangle|\Phi_{E_1E_2}^-\rangle +
|\Phi_{AB}^-\rangle|\Phi_{E_1E_2}^+\rangle +
|\Psi_{AB}^+\rangle|\Psi_{E_1E_2}^-\rangle +
|\Psi_{AB}^-\rangle|\Psi_{E_1E_2}^+\rangle
\right)$\\
\hline
$|-\rangle, |-\rangle$ &
$\frac{1}{2}\left(
|\Phi_{AB}^+\rangle|\Phi_{E_1E_2}^+\rangle +
|\Phi_{AB}^-\rangle|\Phi_{E_1E_2}^-\rangle -
|\Psi_{AB}^+\rangle|\Psi_{E_1E_2}^+\rangle -
|\Psi_{AB}^-\rangle|\Psi_{E_1E_2}^-\rangle
\right)$\\
\hline
\end{tabular}
\end{center}
\caption{State with Eve after the CNOT attack.}
\label{tab1}
\end{table*}
After Bob and Alice publicly declares their bases, if that is $Z$, then 
Eve obtains all the information without introducing any error by measuring in
computational basis $|00\rangle, |01\rangle, |10\rangle, |11\rangle$.
If the basis is $X$, then Eve's interaction introduces error at the rate of 
$\frac{1}{2}$, but Eve does not obtain any information about the secret bits. 
In these cases, if Eve measures $|\Phi_{E_1E_2}^+\rangle$ or
$|\Psi_{E_1E_2}^+\rangle$ then she knows that no disturbance has been 
introduced. If Eve measures $|\Phi_{E_1E_2}^-\rangle$ or
$|\Psi_{E_1E_2}^-\rangle$ then she understands that error has been 
introduced, i.e., Bob and Alice will land into a complement bit value 
at this location of the secret key.
To summarize, we have the following situation.
\begin{center}
\begin{tabular}{|c|c|c|c|c|c|c|}
\hline
Basis & Operation & $BER_{AB}$ & $P_E$ & $\pi_E$\\
(Alice, Bob) & by Eve & & & \\
\hline
$Z$ & CNOT & 0 & 1 & 1\\
\hline
$X$ & CNOT & 0.5 & 0.5 & 1\\
\hline
\end{tabular}
\end{center}

\subsection{The symmetric version}
\label{three2}
As we have seen in the previous section, the third-party CNOT attack does
not introduce any error in $Z$ basis, but induces errors in half of the cases
in $X$ basis. Thus this eavesdropping scenario is asymmetric. To provide
a symmetric scenario, we make the following modification. 

Let $H$ be the Hadamard gate, $I$ be the identity gate
(both works on a single qubit) and $C$ be the CNOT gate (that works on
two qubits). Let us define
$$P_u = (H \otimes I)^u C (H \otimes I)^u \mbox { for } u = 0, 1,$$ i.e., 
$P_0 = C$ and $P_1 = (H \otimes I) C (H \otimes I)$.
One can check that 
\begin{center}
\begin{tabular}{ll}
$P_0(|00\rangle) = |00\rangle$, & $P_0(|10\rangle) = |11\rangle$,\\
$P_0(|+0\rangle) = \frac{|++\rangle + |--\rangle}{\sqrt{2}}$, &
$P_0(|-0\rangle) = \frac{|+-\rangle + |-+\rangle}{\sqrt{2}}$,\\
$P_1(|00\rangle) = \frac{|0+\rangle + |1-\rangle}{\sqrt{2}}$, &
$P_1(|10\rangle) = \frac{|0-\rangle + |1+\rangle}{\sqrt{2}}$,\\
$P_1(|+0\rangle) = |+0\rangle$, & $P_1(|-0\rangle) = |-1\rangle$.
\end{tabular}
\end{center}
Eve applies either $P_0$ or
$P_1$ based on the outcome of an unbiased coin toss. The case of applying
$P_0$ (CNOT) for each of Bob and Alice has been described in previous section.

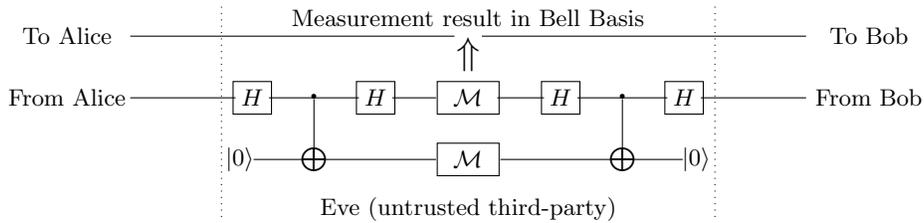
\begin{figure*}[htb]
\begin{center}
\begin{tikzpicture}[scale=0.82]
\node [inner sep=0pt,outer sep=0pt](p1) at (1,1) {};
\node [inner sep=0pt,outer sep=0pt](p2) at (4,1) {};
\node [inner sep=0pt,outer sep=0pt](p5) at (2,1) {};
\node [inner sep=0pt,outer sep=0pt](p6) at (2,2) {};
\draw [-] (p1) -- (p2);
\draw [-] (p5) -- (p6);
\node [inner sep=0pt,outer sep=0pt] (xor1) at (2,1) {$\bigoplus$};
\node [inner sep=0pt,outer sep=0pt] (dot1) at (2,2) {{\LARGE$\cdot$}};

\node [inner sep=0pt,outer sep=0pt](p3) at (-1,2) {};
\node [inner sep=0pt,outer sep=0pt](p4) at (0.75,2) {};
\draw [-] (p3) -- (p4);
\node [inner sep=0pt,outer sep=0pt](p3) at (1.25,2) {};
\node [inner sep=0pt,outer sep=0pt](p4) at (2.75,2) {};
\draw [-] (p3) -- (p4);
\node [inner sep=0pt,outer sep=0pt](p3) at (3.25,2) {};
\node [inner sep=0pt,outer sep=0pt](p4) at (4,2) {};
\draw [-] (p3) -- (p4);
\node [inner sep=0pt,outer sep=0pt](p3) at (4.95,2) {};
\node [inner sep=0pt,outer sep=0pt](p4) at (5.75,2) {};
\draw [-] (p3) -- (p4);
\node [inner sep=0pt,outer sep=0pt](p3) at (6.25,2) {};
\node [inner sep=0pt,outer sep=0pt](p4) at (7.75,2) {};
\draw [-] (p3) -- (p4);
\node [inner sep=0pt,outer sep=0pt](p3) at (8.25,2) {};
\node [inner sep=0pt,outer sep=0pt](p4) at (10,2) {};
\draw [-] (p3) -- (p4);

\node [inner sep=0pt,outer sep=0pt](p1) at (5,1) {};
\node [inner sep=0pt,outer sep=0pt](p2) at (8,1) {};
\node [inner sep=0pt,outer sep=0pt](p5) at (7,1) {};
\node [inner sep=0pt,outer sep=0pt](p6) at (7,2) {};
\draw [-] (p1) -- (p2);
\draw [-] (p5) -- (p6);
\node [inner sep=0pt,outer sep=0pt] (xor1) at (7,1) {$\bigoplus$};
\node [inner sep=0pt,outer sep=0pt] (dot1) at (7,2) {{\LARGE$\cdot$}};

\node [inner sep=0pt,outer sep=0pt] (t1) at (-2,2) {From Alice};
\node [inner sep=0pt,outer sep=0pt] (t2) at (0.8,1) {$|0\rangle$};
\node [inner sep=0pt,outer sep=0pt] (t2) at (8.2,1) {$|0\rangle$};
\node [inner sep=0pt,outer sep=0pt] (t2) at (11,2) {From Bob};
\node [inner sep=0pt,outer sep=0pt] (t1) at (-2,3) {To Alice};
\node [inner sep=0pt,outer sep=0pt] (t2) at (11,3) {To Bob};
\node [inner sep=0pt,outer sep=0pt] (t2) at (4.5,0.2) {Eve (untrusted third-party)};
\node [inner sep=0pt,outer sep=0pt] (t2) at (4.5,3.3) {Measurement result in Bell Basis};

\node[rectangle, draw=black, minimum height=2, minimum width=2] (H) at (4.5,1) {$ \ \cal{M} \ $};
\node[rectangle, draw=black, minimum height=2, minimum width=2] (H) at (4.5,2) {$ \ \cal{M} \ $};

\node [inner sep=0pt,outer sep=0pt](p1) at (0.5,3.5) {};
\node [inner sep=0pt,outer sep=0pt](p2) at (0.5,0) {};
\draw [dotted] (p1) -- (p2);
\node [inner sep=0pt,outer sep=0pt](p1) at (8.5,3.5) {};
\node [inner sep=0pt,outer sep=0pt](p2) at (8.5,0) {};
\draw [dotted] (p1) -- (p2);
\node [inner sep=0pt,outer sep=0pt](p1) at (-1,3) {};
\node [inner sep=0pt,outer sep=0pt](p2) at (4.3,3) {};
\draw [-] (p1) -- (p2);
\node [inner sep=0pt,outer sep=0pt](p1) at (10,3) {};
\node [inner sep=0pt,outer sep=0pt](p2) at (4.7,3) {};
\draw [-] (p1) -- (p2);
\node [inner sep=0pt,outer sep=0pt] (t1) at (4.5,2.7) {\Large{$\Uparrow$}};
\node[rectangle, draw=black, minimum height=2, minimum width=2] (H) at (3,2) {$H$};
\node[rectangle, draw=black, minimum height=2, minimum width=2] (H) at (1,2) {$H$};
\node[rectangle, draw=black, minimum height=2, minimum width=2] (H) at (6,2) {$H$};
\node[rectangle, draw=black, minimum height=2, minimum width=2] (H) at (8,2) {$H$};
\end{tikzpicture}
\caption{Hadamard assisted CNOT attack on MDI QKD}\label{fig7}
\end{center}
\end{figure*}

In case, $P_1$ is applied, and both Bob and Alice communicate in $X$ basis
then Eve will be able to obtain the secret bits completely without creating
any disturbance by measuring her target qubits
in computational basis $|00\rangle, |01\rangle, |10\rangle, |11\rangle$. 

However, if Alice and Bob communicate in $Z$ basis and $P_1$ is applied, then 
Eve will only be able to predict the secret bit as in the case of random 
guess (i.e., with probability $\frac{1}{2}$), though she will be able to 
exactly identify whether error has been introduced. This case is similar to
the one where Alice and Bob communicate in $X$ basis and $P_0$, i.e., CNOT
is applied. The different cases are explained as follows in Table~\ref{tab2}
\begin{table*}[htb]
\begin{center}
\begin{tabular}{|c|c|}
\hline
Alice, Bob & Eve (after Hadamard assisted CNOT attack, i.e., with $P_1$)\\
\hline
$|0\rangle, |0\rangle$ &
$\frac{1}{2}\left(
|\Phi_{AB}^+\rangle|\Phi_{E_1E_2}^+\rangle +
|\Phi_{AB}^-\rangle|\Psi_{E_1E_2}^+\rangle +
|\Psi_{AB}^+\rangle|\Phi_{E_1E_2}^-\rangle -
|\Psi_{AB}^-\rangle|\Psi_{E_1E_2}^-\rangle
\right)$\\

\hline

$|0\rangle, |1\rangle$ &

$\frac{1}{2}\left(
|\Phi_{AB}^+\rangle|\Phi_{E_1E_2}^-\rangle -
|\Phi_{AB}^-\rangle|\Psi_{E_1E_2}^-\rangle +
|\Psi_{AB}^+\rangle|\Phi_{E_1E_2}^+\rangle +
|\Psi_{AB}^-\rangle|\Psi_{E_1E_2}^+\rangle
\right)$\\

\hline

$|1\rangle, |0\rangle$ &
$\frac{1}{2}\left(
|\Phi_{AB}^+\rangle|\Phi_{E_1E_2}^-\rangle +
|\Phi_{AB}^-\rangle|\Psi_{E_1E_2}^-\rangle +
|\Psi_{AB}^+\rangle|\Phi_{E_1E_2}^+\rangle -
|\Psi_{AB}^-\rangle|\Psi_{E_1E_2}^+\rangle
\right)$\\

\hline

$|1\rangle, |1\rangle$ &
$\frac{1}{2}\left(
|\Phi_{AB}^+\rangle|\Phi_{E_1E_2}^+\rangle -
|\Phi_{AB}^-\rangle|\Psi_{E_1E_2}^+\rangle +
|\Psi_{AB}^+\rangle|\Phi_{E_1E_2}^-\rangle +
|\Psi_{AB}^-\rangle|\Psi_{E_1E_2}^-\rangle
\right)$\\
\hline
\end{tabular}
\end{center}
\caption{State with Eve after the Hadamard assisted CNOT attack.}
\label{tab2}
\end{table*}
Thus, when $P_0$ and $P_1$ are used randomly with probability $\frac{1}{2}$ 
in each case, we have the following outcomes.
\begin{center}
\begin{tabular}{|c|c|c|c|c|c|c|}
\hline
Basis & Operation & $BER_{AB}$ & $P_E$ & $\pi_E$\\
(Alice, Bob) & by Eve & & & \\
\hline
$Z$ & $P_0$ & 0 & 1 & 1\\
\hline
$X$ & $P_0$ & 0.5 & 0.5 & 1\\
\hline
$Z$ & $P_1$ & 0.5 & 0.5 & 1\\
\hline
$X$ & $P_1$ & 0 & 1 & 1\\
\hline
\end{tabular}
\end{center}
The complete algorithm for Hadamard assisted CNOT attack is as follows.
\begin{enumerate}
\item Eve applies either $P_0$ or $P_1$ on the qubits 
$|\mu_A\rangle$ and $|\mu_B\rangle$ (communicated by Alice and Bob to Eve) and 
$|0\rangle$ (ancilla supplied by Eve) for both the cases.

\item The two-qubit state (outputs corresponding to $|\mu_A\rangle, 
|\mu_B\rangle$) is measured in Bell basis and the result is communicated to 
Alice and Bob. Further, both the outputs corresponding to the $|0\rangle$ 
qubits (the target ones) are kept with Eve.

\item After the public discussion between Alice and Bob, Eve comes to know
about the cases where Alice and Bob communicated in the same basis. 
The cases where Bob and Alice
communicated in different bases are in any case discarded.

\item If Alice and Bob both communicated qubits in $Z$ (respectively $X$) basis 
and Eve applied $P_0$ (respectively $P_1$), then Eve obtains 
the corresponding secret bit correctly without introducing any error
by measuring the pair of qubits in computational basis.

\item If Alice and Bob both communicated qubits in $Z$ (respectively $X$) basis 
and Eve applied $P_1$ (respectively $P_0$), then Eve can only guess about the
communicated bit with probability $\frac{1}{2}$ (i.e., no information better
than the random guess) inducing a bit error with probability $\frac{1}{2}$. 

In such cases, Eve measures her qubits in Bell basis and if the measurement 
output is $|\Phi_{E_1E_2}^+\rangle$ or $|\Psi_{E_1E_2}^+\rangle$
(respectively $|\Phi_{E_1E_2}^-\rangle$ or $|\Psi_{E_1E_2}^-\rangle$)
then Eve knows that error has not been 
(respectively has been) introduced in the communication.
\end{enumerate}

As Alice and Bob settle on either $Z$ or $X$ basis equally likely, and Eve
also applies $P_0$ or $P_1$ based on the outcome of an unbiased coin,
the error rate in both $Z$ and $X$ basis will be equal. Thus the attack
is a symmetric one. On an average, $BER_{AB} = \frac{1}{4}$, 
$P_E = \frac{3}{4}$ and $\pi_E = 1$.

Moreover, the eavesdropping by Eve may be induced in a portion of the 
communicated bits instead of all, say a proportion $\zeta$. This is due
to the fact that if Alice and Bob notice a channel noise more than some 
threshold value, then they will abort the protocol. In such a case, Eve
will be able to guess expected $\frac{\zeta}{2}$ proportion of bits with 
probability 1. For the remaining bits, though she will not gain anything
other than the random guess, she will be able to know whether error has been
induced during the communication between Alice and Bob. Thus, on an average, 
$BER_{AB} = \frac{\zeta}{4}$, $P_E = \frac{3\zeta}{4}$ and $\pi_E = \zeta$.
Due to the symmetric nature of this eavesdropping strategy, Alice and Bob 
would not be able to distinguish this eavesdropping from channel noise.

\section{Conclusion}
In this letter, we have considered how CNOT kind of attack can be mounted on
a recently proposed variant of BB84, which is referred as Measurement Device
Independent Quantum Key Distribution (MDI QKD) protocol~\cite{side2}. Though
our analysis is in the direction of CNOT attack on BB84~\cite{bb84}, 
it requires a different approach by the third-party to execute the attack
exploiting entanglement swapping.  
Through this kind of eavesdropping, Eve will exactly obtain around half of 
the secret bits communicated between Alice and Bob. For the rest of the bits, 
Eve will only be able to predict the bit values as in random guess. However, 
she will certainly find out whether her interaction induced an error
between Alice and Bob.

\end{document}